\documentclass{aa}

\usepackage{graphicx}
\usepackage{longtable}
\usepackage{txfonts}
\usepackage{lscape}
\usepackage{threeparttable}

\begin{document}

\title{CD-62$\degr$1346:  an extreme halo or hypervelocity CH~star?
\thanks{Based on observations made with the 2.2m telescope at the European 
Southern Observatory (La Silla, Chile).}}

  \author{C.B. Pereira\inst{1}, E. Jilinski\inst{1,2,3}, N.A.~ Drake\inst{1,4}, 
D.B. de Castro\inst{1}, V.G.~Ortega\inst{1}, C.Chavero\inst{1,5} \& F. Roig\inst{1}}

   \offprints{C.B. Pereira}

\institute{Observat\'orio Nacional, Rua Jos\'e Cristino 77, 
CEP 20921-400, S\~ao Crist\'ov\~ao, Rio de Janeiro-RJ. Brazil.
    \and Instituto de F\'{\i}sica, Universidade do Estado do Rio de Janeiro, 
Rua S\~ao Francisco Xavier 524, Maracan\~a, 200550-900 Rio de Janeiro-RJ, Brazil
    \and Pulkovo Observatory, Russian Academy of Sciences, 65, Pulkovo, 196140,
St. Petersburg, Russia 
    \and Sobolev Astronomical Institute, St.~Petersburg State University, 
Universitetski pr.~28, St.~Petersburg 198504, Russia.
    \and Observatorio Astr\'onomico de C\'ordoba, Laprida 854, C\'ordoba, 5000, 
Argentina\\
   \email{claudio,jilinski,drake,denise,vladimir,carolina,froig@on.br}}

  \date{Received ; accepted }

\abstract 
{High-velocity halo stars provide important information about the 
properties of the extreme Galactic halo. The study of unbound and bound Population~II 
stars permits us to better estimate the mass of the halo.}
{We carried out a detailed spectroscopic and kinematic study and have 
significantly refined the distance and the evolutionary state of the star.} 
{Its atmospheric parameters, chemical abundances and kinematical properties 
were determined using high-resolution optical spectroscopy and employing the 
local-thermodynamic-equilibrium model atmospheres of Kurucz and the
spectral analysis code {\sc moog}.}
{We found that CD-62$\degr$1346 is a metal-poor ([Fe/H]=$-$1.7) evolved giant star 
with $T_{\rm eff}=5300$~K  and $\log g=1.7$. The star exhibits 
high carbon and s-element abundances typical of CH stars. It is also a lead star.
Our kinematic analysis 
of its 3D space motions shows that this star has a highly eccentric ($e=0.91$) retrograde 
orbit with an apogalactic distance of $\sim 100$~kpc, exceeding by a factor of three 
the distance to the Magellanic Clouds. The star travels with very high velocity 
relative to the Galactocentric reference frame ($V_{\rm GRF}=570$~km\,s$^{-1}$).} 
{CD-62$\degr$1346 is an evolved giant star and not a subgiant star, as 
was considered earlier. Whether it is bound or unbound
to the Galaxy depends on the assumed mass and on the adopted Galactic potential.
We also show that the star HD 5223 is another example of a high-velocity
CH star that exceeds the Galactic escape velocity. Possible origins of these
two high-velocity stars are briefly discussed. CD-62$\degr$1346 and HD 5223 are the 
first red giant stars to join the restricted group of hypervelocity stars.}


   \authorrunning{C.B. Pereira et al.}
   \titlerunning{CD-62$\degr$1346 : extreme halo star}

\maketitle

\section{Introduction}

\par The first work that identified stars moving with extreme
velocities in the Galactic rest frame $V_{\rm GRF}$ (extreme-velocity
stars) was   that of Carney \& Peterson 1988,  and references therein.  These authors
measured  high proper motions and found five
stars with $V_{\rm GRF} \ge 380$~km\,s$^{-1}$, where $V_{\rm
  GRF}=\sqrt{U_{0}^2+V_{0}^2 +W_{0}^2}$, and $U_{0}$, $V_{0}$ and
$W_{0}$ are the star Galactocentric velocity components.  The highest
velocity halo star of their sample moves through the solar
neighborhood with a $V_{\rm GRF}$ of 490~km\,s$^{-1}$.  However, none
of these stars exceeds the Galaxy's escape velocity ($\sim\!
536$~km\,s$^{-1}$ at the distance of the Sun, according to the Galactic
potential of Allen \& Santillan 1991) and are bound to the Galaxy.
The first hypervelocity star ($V_{\rm GRF}=709$~km\,s$^{-1}$), i.e. an
unbound star traveling with a $V_{\rm GRF}$ velocity higher than the
escape velocity, was discovered by Brown et al. (2005):
SDSS~J090745.0+024507, a late B-type main-sequence star at the
distance of $\sim\! 110$~kpc in the Galactic halo.  According to Hills
(1988), hypervelocity stars (HVSs) are a natural consequence of the
presence of a massive black hole in the Galactic center.  One year
later four additional HVSs were discovered, SDSS~J091301.0+305120 and
SDSS~J91759.5+672238 (Brown et al. 2006a) and SDSS~J110557.45+093439.5
and SDSS~J113312.12+010824 (Brown et al. 2006b), traveling with $V_{\rm
  GRF}$ at least $+558\! \pm\!12$, $+638\pm\!  12$, $+508\pm\! 12$,
and $+418 \pm\!  10$~km\,s$^{-1}$, respectively. Assuming that these
HVSs are B8 main-sequence stars, Brown and collaborators  estimated their
heliocentric distances to be 75, 55, 75, and 55\,kpc, respectively.

\par Up to now 17 HVS have been found (Brown et al. 2009) and most of the
known HVSs are early-type main-sequence stars. Only a very limited
number of evolved HVSs was discovered: US~708, a low-mass hot subdwarf
star of spectral type sdO (Hirsch et al. 2005) and SDSS
\,J153935.67$+$023909.8, a Population~II horizontal branch star with
$V_{\rm GRF}\sim\! 700$ km\,s$^{-1}$, one of the fastest halo stars
known (Przybilla et al. 2010).  Following the discovery of US 708,
Tillich et al. (2011) searched for other hot subdwarfs and found 10 new
subdwarfs HVS candidates, which are SDSS J121150.27$+$143716.2, even
faster than J153935.67$+$023909.8, with $V_{\rm GRF}\sim\! 713$
km\,$^{-1}$.  Kinematic studies of the HVSs are based mainly on their
radial velocities (RVs) alone. Proper motions can be measured only for
a few HVSs (Tilich et al. 2009, 2011).  Field halo stars and  globular clusters have been used to estimate the mass of
the halo.  However, only the objects with the most extreme velocities
may provide tight constraints for the mass estimates (Sakamoto et al.
2003; Smith et al.  2007). Therefore, the study of additional high-velocity
stars in the halo and consideration of stars other than B/BHB stars is
of the utmost importance.

\par In this work we report the discovery of a CH star traveling with
$V_{\rm GRF}=577$~km\,s$^{-1}$ close to the local escape velocity
$v_{\rm esc}$=664.5~km\,s$^{-1}$ at the star distance, determined
using the Galactic gravitational potential of Ortega et al. (2002).
As we will show in Section~4.4, CD-62$\degr$1346 belongs to the small
known group of fastest moving stars in the halo, and is one of the
first red giant stars to be candidate for an HVSs group.
CD-62$\degr$1346 was first recognized as a metal-poor
chemically peculiar star by Bond (1970b) and was later
spectroscopically investigated by Luck \& Bond (1991) in an extensive
investigation performed by these authors on CH and barium stars.
CD-62$\degr$1346 was included in the Kapteyn group by Wylie-de Boer et
al. (2010) based on the short star\,-\,Sun distance (300~pc) adopted
by these authors.  CD-62$\degr$1346 was selected as a target to be
observed in the framework of our high-resolution spectroscopy survey
dedicated to analyze a large sample of chemically peculiar barium
stars from MacConnell et al. (1972) and some stars from Luck \& Bond
(1991). Of the 230 surveyed stars, we have already discovered a new CH
subgiant, BD-03$^\circ$3668 (Pereira \& Drake 2011) and  a
sample of metal-rich barium stars (Pereira et al. 2011).  To
understand the origin of an extreme high-velocity star,  it is necessary
to know its detailed abundance pattern.  Hence, a detailed kinematic
and chemical abundance study of the star is needed. In the course of
our analysis we found it interesting to compare the kinematic properties
of other CH stars that are already known in the literature, with those of
CD-62$\degr$1346. As we will  show, we found that another CH star,
HD~5223, has a $V_{\rm GRF}$ of ~713 km\,s$^{-1}$, a velocity similar
to that of one of the fastest known halo stars: SDSSJ153935.67$+$023909.8
(Przybilla et al. 2010) with $V_{\rm GRF}=694$~km\,s$^{-1}$.

\section{Observations}

\par The high-resolution spectrum of CD-62$\degr$1346 analyzed in
this work was obtained with the FEROS (Fiberfed Extended Range Optical
Spectrograph) echelle spectrograph (Kaufer et al. 1999) at the 2.2\,m ESO
telescope at La Silla (Chile) on the night of October 19, 2008.  The FEROS
spectral resolving power is $R=48\,000$, corresponding to 2.2 pixels of
$15\,\mu$m, and the wavelength coverage goes from 3\,800\,{\AA} to
9\,200\,{\AA}.  The nominal signal-to-noise ratio ($S/N$)  was evaluated by measuring the rms
flux fluctuation in selected continuum windows, and the typical value was $S/N
= 100-150$ after one exposure of 1200 secs. The spectra were reduced with the
MIDAS pipeline reduction package consisting of the following standard steps:
CCD bias correction, flat-fielding, spectrum extraction, wavelength
calibration, correction of barycentrer velocity, and spectrum rectification.
Figure~1 shows the spectrum of CD-62$\degr$1346 in the 6130 -- 6170~\AA\ 
region. A redshift of about 2.6~\AA\, due to the radial velocity of 
126.5~km\,s$^{-1}$ is observed.

\section{Analysis and results} 

\subsection{Line selection, equivalent width measurements and oscillator strengths} 

\par The atomic absorption lines selected in the present study are
basically the same as used in previous studies devoted to the analysis of
photospheric abundances of chemically peculiar stars (Pereira \& Drake
2009).  The selected lines are sufficiently unblended to yield
reliable abundances.  Table~1 shows the \ion{Fe}{i} and \ion{Fe}{ii}
lines employed in the analysis, the lower excitation potential
$\chi$(eV) of the transitions, the $\log gf$ values, the measured
equivalent widths $EW_\lambda$ and the derived iron abundances for
each line. The $\log gf$ values for the Fe\,{\sc i} and Fe\,{\sc ii}
lines given in Table~1 were taken from Lambert et al.  (1996).
\addtocounter{table}{1}

\begin{figure}  
 \centering
\includegraphics[width=9.1cm]{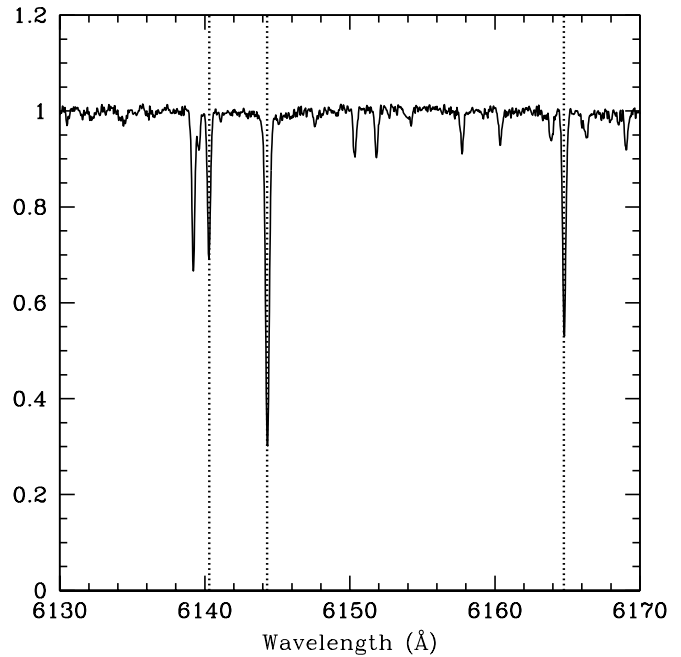}
\caption{Spectrum of CD-62$\degr$1346.
Short dashed vertical lines show the transitions of Fe\,{\sc i} 6137.70, 
Ba\,{\sc ii} 6141.73 and Ca\,{\sc i} 6162.12. The spectral lines are redshifted 
 by about 2.6~\AA\, due to a radial velocity of 
126.5~km\,s$^{-1}$.}
\end{figure}

\subsection{Determination of the atmospheric parameters}  

\par The determination of stellar atmospheric parameters such as effective
temperature ($T_{\rm eff}$), surface gravity ($\log g$),
microturbulence ($\xi$), and [Fe/H] (throughout, we use the notation
[X/H]=log(N(X)/N(H))$_{\star}$\,$-$\,log(N(X)/N(H)$_{\odot}$)) are
prerequisites for determining  photospheric abundances. The
atmospheric parameters were determined using the local thermodynamic
equilibrium (hereafter LTE) model atmospheres of Kurucz (1993) and the
spectral analysis code MOOG (Sneden 1973).

\par The solution of the excitation equilibrium used to derive the
effective temperature ($T_{\rm eff}$) was defined by a zero slope of
the trend between the iron abundances derived from Fe\,{\sc i} lines
and the excitation potential of the measured lines.  The
microturbulent velocity ($\xi$) was found by constraining the
abundance, determined from individual Fe\,{\sc i} lines, to show no
dependence on $W_{\lambda}/{\lambda}$.  The solution thus found is
unique, depending only on the set of Fe\,{\sc i,\sc ii} lines and the
atmospheric model employed. As a by-product this yields the
metallicity [Fe/H] of the star.  The value of log $g$ in Table 3 was
determined by means of the ionization balance assuming
LTE. The final adopted atmospheric parameters are given in Table~3.
Typical uncertainties of $\sigma$($T_{\rm eff}$) = 120\,K,
$\sigma (\log g)\!=\! 0.2$~dex, and $\sigma$($\xi$) =
0.3~km\,s$^{-1}$ were found.  Table~3 also provides previous atmospheric
parameter determinations derived in other studies.

\begin{table*} 
\caption{Atmospheric parameters of CD-62$\degr$1346.}
\begin{tabular}{lccccc}\hline\hline
$T_{\rm eff}$ (K) & $\log g$ (dex) & $[$Fe/H$]$ (dex) & $\xi$ (km\,s$^{-1}$) &
Method ({\rm s/p})$^{a}$ & Reference \\
\hline
5\,300 $\pm$120   & 1.7$\pm$0.2    & $-$1.59$\pm$ 0.08 & 2.1$\pm$0.3   & {\rm s} & 1 \\
5\,150            & 1.92           & -1.39            & 1.9            & {\rm s} & 2 \\
  --              &      --        & -1.56            &                & {\rm p} & 3 \\
5\,230            & 1.6            & -1.56            & 1.2            & {\rm s} & 4 \\
5\,450            & 3.5            & -0.97            & 2.7            & {\rm s} & 5 \\
                  &                & -1.63            &                & {\rm p} & 6 \\
\hline
\end{tabular}

\par a: ({\bf s}pectroscopic/{\bf p}hotometric)
\par 1: This work
\par 2: Wylie-de Boer et al. (2010)
\par 3: Beers et al. (2000)
\par 4: Ryan \& Lambert (1995)
\par 5: Luck \& Bond (1991)
\par 6: Norris et al. (1985)
\end{table*}

\subsection{Abundance analysis} 

\par The abundances of chemical elements were determined with the LTE
model atmosphere techniques.  In brief, equivalent widths are
calculated by integrating through a model atmosphere and are then
compared with the observed equivalent widths. The calculations are
repeated, changing the abundance of the element in question, until a
match is achieved. The current version of the line-synthesis code {\sc
  moog} (Sneden 1973) was used to carry out the calculations.  Table~3
shows the atomic lines used to derive the abundances of the
elements. Atomic parameters for several transitions of Ti, Cr, and Ni
were retrieved from the library of the National Institute of Science
and Technology Atomic Spectra Database (Martin 2002).  
The derived abundances and the C/O ratio are given in Table~4.
\addtocounter{table}{1}

\par The barium abundance was derived using the Ba\,{\sc ii} lines at
$\lambda$ 4554.0, $\lambda$ 4934.1, $\lambda$ 5853,7, and
$\lambda$ 6141.7\AA. The line data  that include hyperfine splitting
were taken from McWilliam (1998).

\par The lead abundance was derived from the Pb\,{\sc i} line at
$\lambda 4057.81$~\AA.  The line data, which include isotopic shifts
and hyperfine splitting, were taken from Van Eck et al. (2003).

\par Carbon, nitrogen, and oxygen abundances were also determined
using the spectrum synthesis technique.  Since the abundances of the
CNO elements are interdependent because of the association of carbon
and oxygen in CO molecules in the atmospheres of cool giants, the CNO
abundance determination procedure was iterated until all 
abundances of these three elements agreed.  The abundances of carbon
and nitrogen were determined using the lines of the CH, CN, and C$_2$
molecules.  The line lists were assembled by ourselves and are the
same as in Drake \& Pereira (2008) and Pereira \& Drake (2009), who
studied the chemically peculiar metal-poor stars HD~104340, HD~206983,
HD~10613, and BD+04$\degr$2466.  Nevertheless, studying the Fehrenbach
\& Duflot star (hereafter Feh-Duf), we have updated the value of the
dissociation potential of the CN molecule (Drake \& Pereira 2011).  In
the above mentioned papers, we used the value $D_0$(CN)$=7.65$~eV
determined by Engleman \& Rouse (1975). More recent studies based on
experimental determinations (Huang et al. 1992; Costes et al.  1990)
as well as on theoretical calculations (Pradhan et al. 1994) showed
that the dissociation energy of the CN molecule is higher, about
7.75~eV, and this value was used in the synthetic spectra
calculations. However, we have to mention that the uncertainty in the
derived nitrogen abundance is high because of the  strong contamination by
telluric H$_2$O lines and  the low S/N ratio in this spectral region.  The
oxygen abundance was inferred from the [O\,{\sc i}] forbidden line at
$\lambda$6300.304~\AA.  In our calculations for this line we used the
oscillator strength $\log gf\!=\! -9.717$ obtained by Allende Prieto et
al. (2001) in their analysis of the solar oxygen abundance.

\begin{table} 
\begin{center}
\caption{Chemical abundances derived for CD-62$\degr$1346 in the scale 
$\log\varepsilon({\rm H}) = 12.0$ and in the notations [X/H] and [X/Fe]. 
The adopted solar abundances are from Grevesse \& Sauval (1998).}
\label{table:4}
\centering
\begin{tabular}{lcccc}\hline\hline
Species & n & $\log\varepsilon$ & [X/H] & [X/Fe]  \\
\hline 
Fe\,{\sc i}  & 52 & 5.93$\pm$0.08 & -1.59 & ---   \\
Fe\,{\sc ii} & 11 & 5.92$\pm$0.07 & -1.60 & ---   \\
C\,(C$_2$)   & syn &7.79$\pm$0.08 & -0.73 & +0.86 \\
N\,(CN)      & syn &7.35:          & -0.57: & +1.02: \\
O\,{\sc i}   &  1 & 7.78          & -1.05 & +0.54 \\
Na\,{\sc i}  &  2 & 4.73          & -1.60 & -0.01 \\
Mg\,{\sc i}  &  5 & 6.68$\pm$0.20 & -0.90 & +0.69 \\
Si\,{\sc i}  &  6 & 6.58$\pm$0.13 & -0.97 & +0.62 \\
Ca\,{\sc i}  & 11 & 5.24$\pm$0.15 & -1.12 & +0.47 \\
Ti\,{\sc i}  & 17 & 3.71$\pm$0.12 & -1.31 & +0.28 \\
Cr\,{\sc i}  &  4 & 4.04$\pm$0.06 & -1.63 & -0.04 \\
Ni\,{\sc i}  &  9 & 4.79$\pm$0.15 & -1.46 & +0.13 \\
Y\,{\sc ii}  &  5 & 1.11$\pm$0.07 & -1.13 & +0.46 \\
Zr\,{\sc ii} &  5 & 1.87$\pm$0.23 & -0.73 & +0.86 \\
Ba\,{\sc ii} &  4 & 2.12$\pm$0.10 & -0.55 & +1.58 \\
La\,{\sc ii} &  5 & 0.76$\pm$0.10 & -0.41 & +1.18 \\
Ce\,{\sc ii} & 13 & 1.23$\pm$0.12 & -0.35 & +1.24 \\
Nd\,{\sc ii} & 17 & 1.16$\pm$0.16 & -0.34 & +1.25 \\
Pb\,{\sc i}  &  1 & 2.41          & $+$0.46 & 2.05\\\hline
\end{tabular}
\par C/O = 1.02
\end{center}
\end{table}

\subsection{Abundance uncertainties}    

\par The uncertainties of the derived abundances for the program stars
are dominated by two main sources: the errors in the stellar parameters and
errors in the equivalent width measurements.

\par The abundance uncertainties owing to errors in the stellar
atmospheric parameters $T_{\rm eff}$, $\log g$, and $\xi$ were
estimated by varying these parameters by their standard errors and
then computing the changes incurred in the element abundances. The
results of these calculations are displayed in columns 2 to 5 of Table
5.

\par The abundance uncertainties owing to errors in the equivalent widths
measurements were computed from an expression provided by Cayrel (1988).
The errors in the equivalent widths are essentially set by the
signal-to-noise ratio and by the resolution of the spectra. In our case, having
$R\approx 50\,000$ and a typical S/N ratio of 150, the expected uncertainties in the
equivalent widths are about 2--3 m{\AA}. 

\par Under the assumption that the errors are independent, they can be
combined quadratically so that the total uncertainty is
\[ \sigma = \sqrt{\sum_{i=1}^{N} \sigma^{2}_{i}.} \]
These final uncertainties are given in the sixth column of Table 5.
The last column gives the observed
abundance dispersion among the lines for those elements with more than three
available lines. Table 5 also shows that neutral elements are fairly sensitive
to temperature variations, while single ionized elements are sensitive
to the variations in log $g$. The uncertainties in microturbulence also
contribute to the compounded errors, especially for strong lines such as the
barium lines. For the elements analyzed via spectrum
synthesis, the same technique was used, by varying $T_{\rm eff}$, $\log g$,
and $\xi$ and then computing the abundance changes introduced by the
variation in these atmospheric parameters. The resulting uncertainties are
also included in Table~5.

\begin{table*}
\caption{Abundance uncertainties for CD-62$\degr$1346. The second
column gives the variation of the abundance  caused by the variation in
$T_{\rm eff}$. The other columns refer to the abundance variations caused by
 the uncertainty in $\log g$, $\xi$, and $W_\lambda$, respectively. The sixth column gives
the compounded r.m.s. uncertainty of the second to fifth columns. The
last column gives the observed abundance dispersion for those elements
whose abundances were derived using more than three lines.}
\label{table:5}
\centering
\begin{tabular}{lcccccc}\hline\hline
Species & $\Delta T_{\rm eff}$ & $\Delta\log g$ & $\Delta\xi$ & $\Delta W_{\lambda}$
& $\left( \sum \sigma^2 \right)^{1/2}$ & $\sigma$$_{\rm obs}$ \\
$_{\rule{0pt}{8pt}}$ & $+120$~K & $+0.2$ & +0.3 & +3 m\AA & & \\
\hline                                     
Fe\,{\sc i}    & +0.12  & +0.00 & -0.05 & +0.08 & 0.15 & 0.08 \\
Fe\,{\sc ii}   & +0.01  & +0.07 & -0.02 & +0.07 & 0.10 & 0.07 \\
Na\,{\sc i}    & +0.06  &  0.00 & -0.01 & +0.09 & 0.11 & ---  \\
Mg\,{\sc i}    & +0.05  &  0.00 & -0.05 & +0.08 & 0.11 & 0.20 \\
Si\,{\sc i}    & +0.03  &  0.00 & -0.01 & +0.09 & 0.10 & 0.13 \\
Ca\,{\sc i}    & +0.08  & -0.01 & -0.07 & +0.05 & 0.12 & 0.15 \\
Ti\,{\sc i}    & +0.12  & -0.01 & -0.04 & +0.06 & 0.14 & 0.12 \\
Cr\,{\sc i}    & +0.12  & -0.01 & -0.04 & +0.06 & 0.14 & 0.06 \\
Ni\,{\sc i}    & +0.10  &  0.00 & -0.01 & +0.09 & 0.13 & 0.15 \\
Y\,{\sc ii}    & +0.06  & -0.07 & -0.09 & +0.06 & 0.14 & 0.07 \\
Zr\,{\sc ii}   & +0.06  & +0.07 & -0.09 & +0.12 & 0.18 & 0.23 \\
Ba\,{\sc ii}   & +0.14  & +0.08 & -0.23 &  ---  & 0.28 & 0.10 \\
La\,{\sc ii}   & +0.08  & +0.07 & -0.02 & +0.07 & 0.13 & 0.10 \\
Ce\,{\sc ii}   & +0.07  & +0.06 & -0.08 & +0.07 & 0.14 & 0.12 \\
Nd\,{\sc ii}   & +0.08  & +0.06 & -0.04 & +0.06 & 0.12 & 0.16 \\
Pb\,{\sc i}    & +0.15  & -0.01 & -0.03 &  ---  & 0.15 & ---  \\
\hline
\end{tabular}
\end{table*}

\section{Discussion}

\subsection{Abundances}  

\par Below we discuss the abundance pattern of CD-62$\degr$1346 by comparing
it with previous studies for some halo population stars and also with
the abundance pattern of chemically peculiar stars with heavy-element
overabundances already reported in the literature. Figure 2 shows the
abundance pattern of CD-62$\degr$1346 analyzed in this work.

\begin{figure} 
   \includegraphics[width=9.1cm]{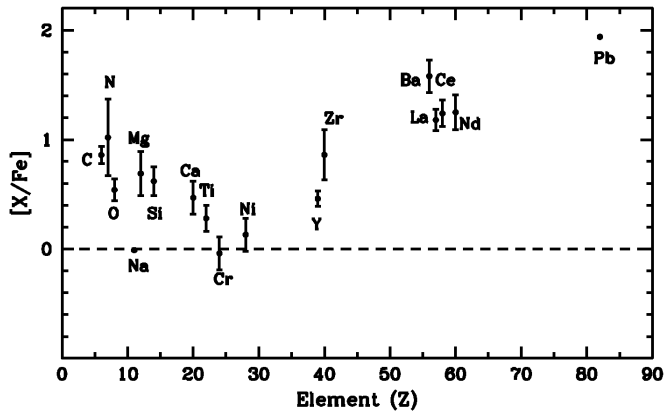}
   \caption{Abundance pattern of CD-62$\degr$1346.
Error bars represent the uncertainty estimates described in the text.}
\end{figure}

\subsubsection{Nitrogen and oxygen}  

\par Abundance surveys of dwarf stars show that there is no trend in
the [N/Fe] $versus$ [Fe/H] ratio, that is, in the metallicity range
$-2.0 <$ [Fe/H] $<+0.3$, [N/Fe] is $\approx 0.0$ (Clegg et al. 1981;
Tomkin \& Lambert 1984; Carbon et al.  1987). In giant stars, the
nuclear processed material, caused by the deepening of its convective
envelope, is brought from the interior to the outer layers of the star,
which changes the surface composition.  As a consequence of the first
dredge-up process, the abundance of $^{12}$C is reduced and the
abundance of nitrogen is enhanced (Lambert 1981).  The
nitrogen-to-iron ratio in CD-62$\degr$1346 is high, [N/Fe]\,=\,1.0,
similar of that of  the mean [N/Fe] ratio observed in the CH stars analyzed by
Vanture (1992b) ([N/Fe]\,=$1.3\pm 0.6$) and also to that in the CH
star BD+04$\degr$2466 analyzed by Pereira \& Drake (2009) with
[N/Fe]\,=$1.1\pm 0.3$. However, this value has to be considered with 
caution because  the $\sim 8000$~\AA, used for the nitrogen
determination, is contaminated by telluric H$_2$O lines and has a
relatively low S/N ratio.
\\
Comparison of the oxygen-to-iron ratio for
CD-62$\degr$1346  with Galactic halo stars indicates that the
[O/Fe] is comparable to that of stars with similar metallicity in the Galaxy
(Nissen et al. 2002) (see Figure~3).

\begin{figure}   
\includegraphics[width=9.1cm]{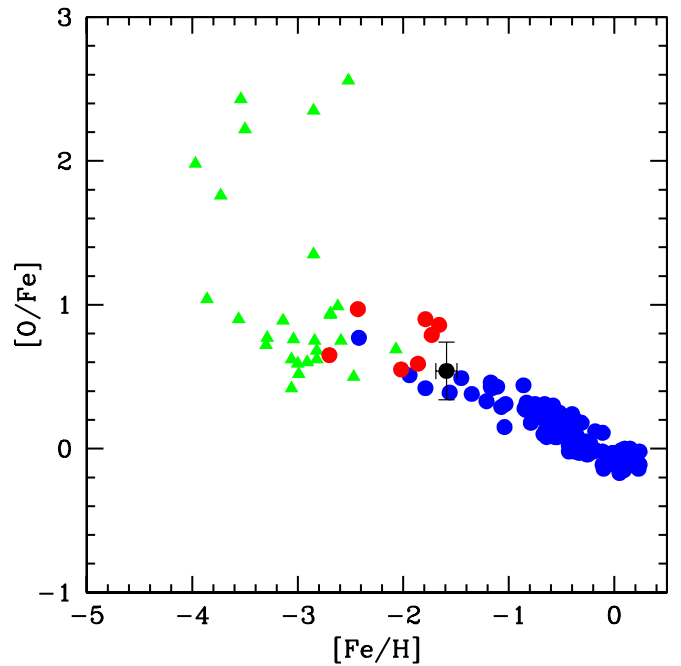}
\caption{[O/Fe] $versus$ [Fe/H] for CD-62$\degr$1346  ({\it black circle}),
disk stars  ({\it blue circles}), CEMP-s stars ({\it red circles}), and 
CEMP-no stars ({\it green triangles}). 
Abundance data for the disk stars are taken  from Edvardsson et al. (1993)
and Nissen et al. (2002). Abundance data for CEMP-no stars are taken from  
Cayrel et al. (2004) and from Masseron et al. (2010) and abundance data
for CEMP-s stars are taken  from Masseron et al. (2010).}
\end{figure}

\subsubsection{Sodium to nickel} 

\par In metal-poor stars the sodium abundance was investigated by
Gratton \& Sneden (1987), McWilliam et al. (1995b), Fulbright (2002),
and Fran\c cois et al. (2003). Between [Fe/H]\,$ =\! -1.0$ and $-3.0$,
the [Na/Fe] ratio is roughly constant and begins to decline at
[Fe/H]\,$ =\! -3.0$ (Cayrel et al. 2004). Our value of $-0.01$ for the
[Na/Fe] ratio at [Fe/H]\,=\,$-1.59$ seems to follow the trend observed
in the metal-poor stars, especially for the metallicity range between
$-2.0\! <$\,[Fe/H]\,$<\! -1.0$ (Fulbright 2002).
 
The mean $\alpha$-element abundance is
[(Mg+Si+Ca+Ti)/4Fe] =\,0.52$\pm$0.18, which is similar for stars
at this metallicity (Carretta et al. 2002).  The chromium and nickel
abundances are expected to follow iron and do indeed with [Cr/Fe]
and [Ni/Fe]$=\! -0.04$ and $-0.13$, respectively.  The [Ni/Fe] ratio
remains close to 0.0 in a metallicity range from $-2.0$ to 0.0
(Jonsell et al. 2005).  The chromium-to-iron ratio also displays no
trend at all for metal-poor stars. CD-62$\degr$1346 has a typical
[Cr/Fe] ratio of a metal-poor star at [Fe/H]$\,\sim\! -1.6$.

\subsubsection{Carbon and s-process elements}  

\par Figure 4 shows the [s/Fe] ratio for CD-62$\degr$1346 analyzed in
this work, where 's' represents the mean of the elements created by
slow neutron capture reactions (s-process): Y, Zr, Ba, La, Ce, and Nd.
This figure also shows the [s/Fe] ratios for the barium stars (giants
and dwarfs), the yellow symbiotic stars, the CH stars and the
carbon-enhanced metal-poor (CEMP) stars. We also show the position of
six CEMP stars that are binaries: CS~22942-019, CS~22948-027,
CS~29497-030, CS~29497-034, CS~22964-161, and HE~0024-2523, the data
of which concerning their carbon and heavy-element (Z $>$ 56)
overabundances and binarity, were taken from recent results:
Preston \& Sneden (2001), Sivarani et al.  (2004), Barbuy et
al. (2005), Lucatello et al. (2003), Thompson et al. (2008), Aoki et
al.  (2002), and Hill et al.  (2000).  As we can see from this figure,
CD-62$\degr$1346 is heavily enhanced in the s-process elements, similar
to other CH stars.

\par Figure 5 shows the [C/Fe] abundance ratio plotted
as a function of the metallicity, given by [Fe/H], for the same objects as in
Figure 4. 
This figure shows carbon overabundance for chemically peculiar
objects that are members of binary systems.
 We note that CD-62$\degr$1346 has a ``normal''
[C/Fe] ratio for a chemically peculiar binary star at this metallicity.

\begin{figure}
\centering 
\includegraphics[width=9.5cm]{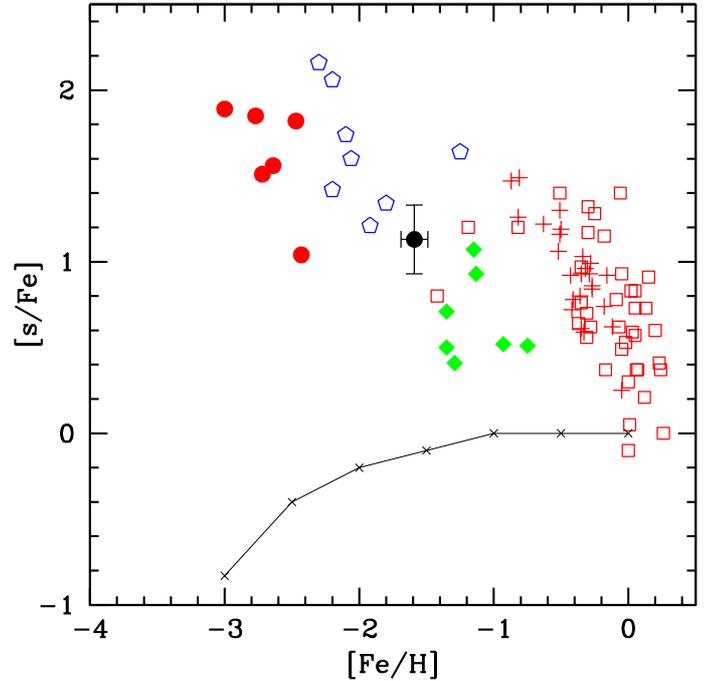}
\caption{Diagram of [s/Fe] {\it versus} [Fe/H]
for several classes of chemically peculiar binary stars.
Barium giants ({\it red open squares}); CH stars ({\it blue open polygons});
subgiant CH stars ({\it red plus signs}); 
yellow-symbiotics ({\it green filled diamonds}); 
CEMP stars that are members  of binary systems ({\it red filled circles}).
The position of CD-62$\degr$1346 is marked by a black filled circle.
Abundance data for barium giant and dwarf stars are taken 
from Allen \& Barbuy (2006), Antipova et al. (2004), Liang et al. (2003), 
North et al. (1994), Smith et al. (1993), 
Drake \& Pereira (2008);
Pereira (2005), Pereira \& Drake (2009,2011); CH stars  are taken from Vanture (1992b) 
and Van Eck et al. (2003); yellow symbiotics are taken from Smith et al. (1996,1997),
and Pereira \& Roig (2009). The solid line is the mean 
$<$[s/Fe]$>$ for field stars (Gratton \& Sneden 1994; Ryan et al. 1996; 
Fran\c cois et al. 2003).}
\end{figure}

\begin{figure} 
\includegraphics[width=9.5cm]{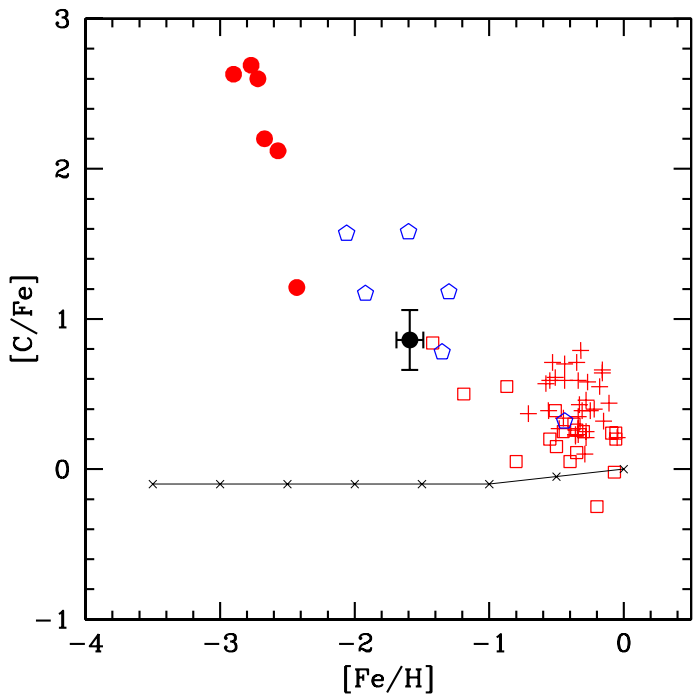}
\caption{Diagram of [C/Fe] {\it versus} [Fe/H]
for several classes of chemically peculiar binary stars and post-AGB stars.
Symbols have the same meaning as in Figure~4. The solid line is the {\sl mean} 
$\langle {\rm [C/Fe]} \rangle$ for field stars, taken from Masseron et
al. (2006).}
\end{figure}

\par The lead abundance was determined using a spectral synthesis
technique.  Figure~6 shows the observed and the synthetic spectra of 
CD-62$\degr$1346 around Pb\,{\sc i} $\lambda$4058~\AA. In Figure~7, we
show the [Pb/Ce] ratio as a function of metallicity [Fe/H]. In this
figure, we show CD-62$\degr$1346 (the filled black circle at
[Fe/H]$\,=\,-1.59$), the CH stars analyzed by Van Eck et al.  (2003),
CEMP stars that are members of binary systems, and the barium giants
and dwarfs from Allen \& Barbuy (2006).  The solid line represents the
prediction from the standard partial mixing (PM) model as given by
Goriely \& Mowlavi (2000). The position of  CD-62$\degr$1346 in this
diagram follows the same trend as other barium dwarfs and giants
investigated so far.  For the [Pb/Ce] ratio the proton-mixing scenario
confirms the trend seen in several CH stars with high lead abundances
(Van Eck et al.  2003; Goriely \& Mowlavi 2000).

\begin{figure} 
\includegraphics[width=9.1cm]{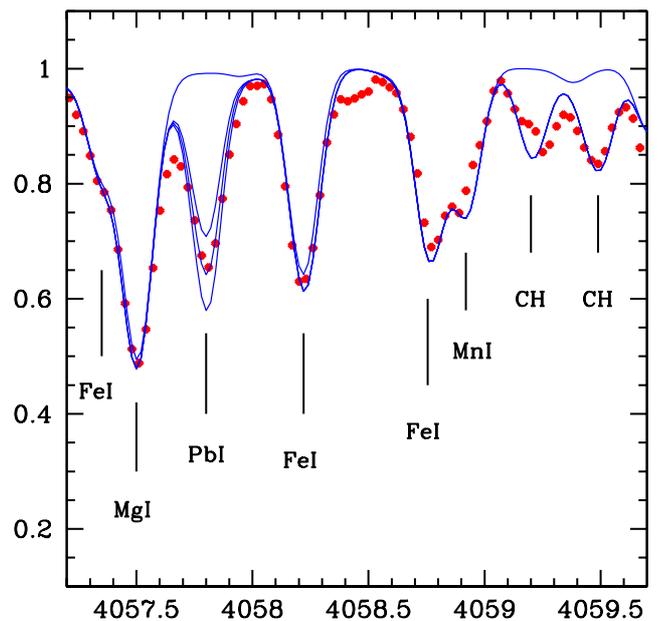}
\caption{Observed (dots) and synthetic (lines) spectra of 
CD-62$\degr$1346 in the region around the Pb\,{\sc i} line at 
$\lambda4057.8$~\AA. 
The synthetic spectra are shown 
for lead abundances of $\log\varepsilon({\rm Pb})=2.21$, 2.41, and 2.61.
The upper line shows a synthesis without contribution from the 
 Pb\,{\sc i} and CH lines.}
\end{figure}

\begin{figure} 
\includegraphics[width=9.1cm]{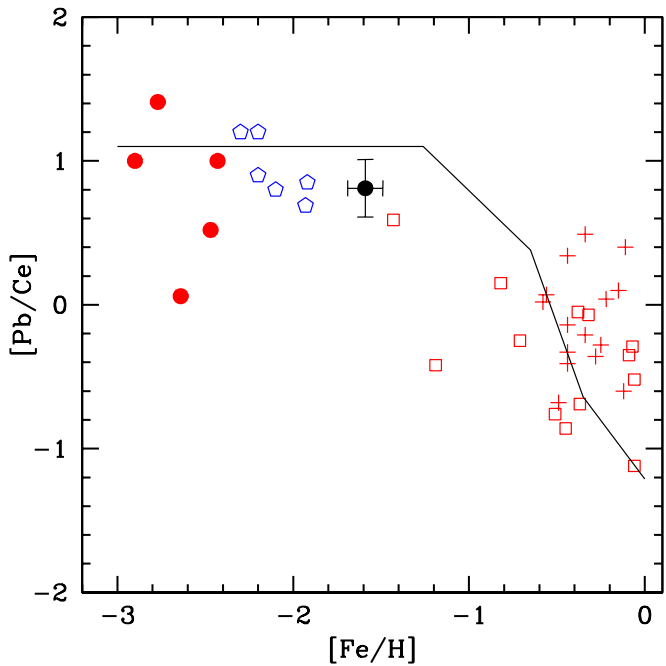}
\caption{Behavior of the [Pb/Ce] ratio with metallicity. Symbols have 
the same meaning as in Figure~4. The solid line represents the prediction 
from the standard partial mixing (PM) model as given by Goriely \& Mowlavi 
(2000).   We considered only those CEMP-s stars	 that have already been 
proven to be binary systems.}
\end{figure}

\par To summarize this section, we find that CD-62$\degr$1346 has a
similar abundance pattern compared to other CH stars, that is,
C/O\,$>$\,1.0, low metallicity, s-process enrichment (it is also a
``lead star'') and a [$\alpha$/Fe] ratio similar to halo stars.
As do a majority of CH stars, CD-62$\degr$1346 has a high radial velocity.
\\
Although the abundance pattern of CH
stars has been well investigated in the last years and, they were also
found to be lead stars (Van Eck et al. 2003), their kinematical
properties were poorly investigated in the literature.  In the next
sections we will explore the results obtained from the kinematical
analysis with the help of the spectroscopic distance and radial
velocity of CD-62$\degr$1346 and compare its space velocity components
with other CH and also some metal-poor barium stars.

\subsection{Distance and luminosity of CD-62$\degr$1346 }

\par The relation between the distance to the Sun $r$, temperature,
gravity, mass, $V$ magnitude, and the interstellar absorption
($A_{V}$) is given by

\begin{eqnarray}
\log r\: ({\rm kpc}) & = & \frac{1}{2}\left(\log \frac{M_{\star}}{M_{\odot}}
+ 0.4\left(V-A_{\rm V}+BC\right) \right. \nonumber \\
& & \left. {{\,}\atop{\,}} + 4\log T_{\rm eff} - \log g - 16.5\right)
\end{eqnarray}

For the derived temperature and the visual magnitude, $V$\,=\,9.85 
(http://simbad.u-strasbg.fr), Equation 1 can be written, as

\begin{equation}
\log r\: ({\rm kpc}) = \frac{1}{2}\left(\log \frac{M_{\star}}{M_{\odot}}\,+\,
2.21- \log g\right).
\end{equation}

\par In these calculations we used $M_{\rm bol\odot}\! =\! +4.74$
(Bessel 1998) and $BC\!=\!-0.19$ (Alonso et al. 1999).  The
interstellar absorption was estimated from the relation between the
equivalent width of the Na\,D$_2$ interstellar line and the color excess
$E(B-V)$ given in Munari \& Zwitter (1997).  For the equivalent width
of 67\,m\AA\, we found a color excess of 0.04.  Finally, for the value
of log $g$\,=\,1.7, and adopting the mass
$M_{*}\,=\,(0.8\pm0.2)\,M_{\sun}$, the most likely value for the mass
of CH stars (McClure \& Woodsworth 1990), we obtain a distance 
$r= 1.6\!\pm\!0.5$~kpc.

\par  The estimation of the distance of CD-62$\degr$1346 deserves
discussion.  A Hipparcos parallax was measured for this star: $\pi =
5.47\pm 1.72$ (van Leeuwen 2007) corresponding to the distance
$d={183^{+83}}\!\!\!\!\!\!\!_{-44}$~pc.  However, this parallax and
the visual magnitude $V=9.85$ correspond to
${M_V=3.4^{+0.6}}\!\!\!\!\!\!\!\!_{-0.8}$ placing CD-62$\degr$1346 at
the beginning of the subgiant branch where the star has to have high
$\log g$ value, incompatible with what we obtained in our spectroscopic
study.  A similar disagreement between the Hipparcos parallax and the
evolutionary state of a star appeared also in the analysis of the
metal-poor halo star BD+17$\degr$3248 performed by Cowan et al. (2002)
forcing the authors to disregard the Hipparcos parallax for
this star. For the same reason we also  disregard the
Hipparcos parallax for CD-62$\degr$1346.  

\par CD-62$\degr$1346 was classified for the first time as a CH
subgiant star by Bond (1974). In his paper, introducing a new class of
peculiar stars - the CH subgiant stars -, luminosity estimations were
based on the reduced proper motions, which for
 the high proper motion star, CD-62$\degr$1346, would result in 
underestimating its luminosity.  Later the photometric estimation
of the absolute magnitude of CD-62$\degr$1346 was carried out by
Norris et al. (1985) based on the $BVRI$ and DDO photometry, which
resulted in a distance of 470~pc. However, the C(3842) and C(4245)
indexes of the DDO system are sensitive to the intensities of CN and
CH molecular bands, because they include the violet CN band and CH G-band,
respectively.  As we have shown in Section~3.3, CD-62$\degr$1346
is a CH star enriched in carbon ([C/Fe]=+0.86). This may explain the
discrepancy in the luminosity found by photometric analysis and by our
spectroscopic determination.

\par Luck \& Bond (1991) used for the analysis
of CD-62$\degr$1346 photographic image-tube spectrogram with the
dispersion 4.6~\AA/mm.  A relatively low spectral resolution
($\sim$0.2~\AA) may explain the higher metallicity ([Fe/H]=--0.97) derived
by these authors.  The higher value of metallicity derived by
means of neutral iron lines might, in turn, result in the high value
of $\log g$ derived by forcing the \ion{Fe}{ii} lines to yield the
same iron abundance.  A more recent analyses based on high-resolution
spectra gave a lower surface gravity and lower metallicity (Ryan \&
Lambert 1995; Wylie-de Boer et al. 2010) (see Table~2).

\par Beers et al. (2000), using the analytical relation between $M_{V}$ and
$(B-V)_{0}$, estimated a distance of 300~pc, a value that was later
adopted by Wylie-de Boer et al. (2010). At this distance,
CD-62$\degr$1346 would have $\log g = 3.2$. However, performing
spectroscopic analysis of this star, Wylie-de Boer et al. (2010) found
a value of $\log g = 1.92$, which is similar to our result.

\par Based on the distance and the interstellar absorption 
determined  above, we estimated the bolometric magnitude and luminosity as
$M_{\rm bol}{}\!=-1.5\!\pm0.7$ and $\log L/L_\odot = 2.5\pm0.2$. Since
CD-62$\degr$1346 displays properties of a halo star (high radial
velocity and high galactic latitude, $b\!=\!  -39.5\degr$), and is
also s-process enriched (Section 4.1.3), it is a CH star and hence a
binary star. Indeed, according to Table~IV of Hartwick \& Cowley
(1985), CH stars have $M_V$ values between $-0.25$ and $-2.2$.
CD-62$\degr$1346 with $M_V\! =\! -1.3$ will be then another CH star,
as we have already concluded in Section~4.1.

\subsection{Radial velocity and proper motions}

\par  Tables 6 and 7 show  all measurements of the radial
velocity and proper motions of CD-62$\degr$1346 available in the
literature.  The proper motions of CD-62$^\circ $1346 seem to be well 
determined, since all measurements give consistent values.

\par  The small variations of  of the observed radial velocity values 
 seen in Table~6 may be the result of the orbit
orientation and the distance between the components of the binary
system.  The first radial velocity measurement was made by Hesser \&
Harris (1979)  and yielded 119~km\,s$^{-1}$. Later
measurements made by Norris et al. (1985), Schuster et al. (2006) and
recently by RAVE (Siebert et al. 2011) gave consistent RV
values. 

\begin{table} 
\caption{References for radial velocity for CD-62$\degr$1346.}
\begin{tabular}{lc}
\hline
Reference & $v_{\rm rad}$ \\
\hline
This work & 126.5$\pm$0.4 \\
Siebert et al. (2011)  & 127.1$\pm$1.1 \\
Schuster et al. (2006) & 123.0$\pm$7.0 \\
Norris et al. (1985) & 127.0 \\
Hesser \& Harris (1979) & 119.0\\\hline
\end{tabular}
\end{table}

\begin{table*} 
\caption{References for proper motions for CD-62$\degr$1346}
\begin{tabular}{ccc}
\hline
                 Catalog                       &   pmRA  &  pmDEC  \\
                                                  &  mas\,yr$^{-1}$ &  mas\,yr$^{-1}$ \\
\hline
Sydney Southern Star Catalog (King \& Lomb 1983)  &  -34.1 & -118 \\
Astrographic Catalog of Reference Stars (ACRS) (Corbin et al. 1991)& -11.1 & -103.1 \\  
Positions and Proper Motions - South (Bastian \& R\"oser 1993) & -22.7 & -111 \\
FOCAT-S Catalog (Bystrov et al. 1994)         & -25    & -111 \\
The Hipparcos and Tycho Catalogs (Perryman 1997)   & -14.97 & -102.19 \\
The ACT Reference Catalog (Urban et al. 1998)    & -13.8 & -107.4 \\
The Tycho Reference Catalog (H\o g et al. 1998) & -14.3 & -104.3 \\ 
The Tycho-2 Catalog (H\o g et al. 2000)         & -16.1 & -103.0 \\
All-sky Compiled Catalog of 2.5 million stars & -14.98 & -103.30 \\
(Kharchenko \& R\"oser 2009)                     &        &          \\
The USNO-B1.0 Catalog (Monet et al. 2003) &  -18 & -104 \\
Hipparcos, the New Reduction (van Leeuwen, 2007) & -15.18 & -102.96 \\
PPMX Catalog of positions and proper motions (R\"oser et al. 2008)& -15.23 & -102.47 \\
UCAC3 Catalog (Zacharias et al. 2009) & -16.0 & -102.0 \\
RAVE 3rd data release (Siebert et al. 2011) & -17.3 & -105.2 \\
SPM4 (Girard et al. 2011)  & -15.79  & -101.87 \\\hline
\end{tabular}
\end{table*}

\subsection{Kinematics}

\par At a distance of 1.6~kpc from the Sun and a mass of $0.8\,
M_\odot$, for CD-62$\degr 1346$, we calculated the heliocentric space
velocity components $U_{0}\!=\!-39.8$~km\,s$^{-1}$,
$V_{0}\!=\!-773.9$~km\,s$^{-1}$ and $W_{0}\!=\!171.2$~km\,s$^{-1}$,
where $U$ is positive toward the Galactic center $(l=0\degr,\;
b=0\degr)$, $V$ is positive in the direction of Galactic rotation
$(l=90\degr,\; b=0\degr)$ and $W$ is positive toward the North
Galactic Pole $(b=90\degr)$.  The algorithm of Johnson \& Soderblom
(1987) was employed in the calculation using Hipparcos proper motions
$\mu$$_{\alpha}$\,cos\,$\delta$\,=\,$-$14.97 and
$\mu$$_{\delta}$\,=\,$-$102.19 (Perryman 1997) and the radial velocity
derived in this study (Table 6).  The transformation from the
heliocentric to the Galactocentric reference frame system (GRF) was
performed using the peculiar solar motion of Co\c skuno\u glu et al. (2011),
the Galactocentric solar distance of 8.5~kpc and Local Standard of
Rest (LSR) rotation velocity relative to the GRF of 220~km\,s$^{-1}$.
The modulus of the star velocity in this reference system $V_{\rm
  GRF}$ can then be computed, giving 569.8~km\,s$^{-1}$.  The high
heliocentric azimuthal velocity component $V$ already suggests that   CD-62$\degr 1346$ may be unbound.  To check this, we
compared the calculated $V_{\rm GRF}$ with the escape velocity at the
star distance.  Using the Galactic gravitational potential described
in Ortega et al. (2002), we found $V_{\rm esc} =664.5$~km\,s$^{-1}$ for
the escape velocity.  So CD-62$\degr 1346$ would be a bound object.
We note, however, that the distances from formula~(2) depend on the adopted
mass of the star.  It is then interesting to see whether the condition
for the star to be bound can change with the distances.  In Table~8 we
show the results for several adopted masses and distances.  For the
mass of $1\,M_\odot$ and a distance of 1.8~kpc, which are within the
uncertainties, CD-62$\degr 1346$ is already becoming unbound.  What
 if other potentials of the Galaxy were used?  We note that
CD-62$\degr 1346$ would be unbound in the potential of Allen \&
Santillan (1991) already at the adopted distance of 1.6~kpc.  If we
used the potential of Johnston et al. (1996), which gives for the escape
velocity $V_{\rm esc} =539$~km\,s$^{-1}$ at distances 1.6 up to 2~kpc,
CD-62$\degr 1346$ would  be an unbound object.  In Section 4.4.2 we show
the orbit of this star calculated with the Ortega et al. (2002)
potential; it probes the Galactic halo up to 100~kpc.  At this
distance the halo-enclosed mass is $6.43\times 10^{11}\, M_\odot$, the
potential by Allen \& Santillan (1991) gives $8\times 10^{11}\,
M_\odot$, while the potential by Johnston et al. (1996) gives
$3.8\times 10^{11}\, M_\odot$.

\begin{table*} 
\caption{Mass, distance, total Galactic reference frame ($V_{\rm GRF}$) 
and the escape velocity calculated for the different stellar masses.}
\begin{tabular}{cccc}\hline\hline
 Mass & distance & $V_{\rm GRF}$ & $v_{\rm esc}$ \\ 
 $M/M_\odot$ & kpc &  km\,s$^{-1}$ & km\,s$^{-1}$ \\
\hline 
0.5 & 1.3  &  424.2 & 664.1 \\
0.6 & 1.4  &  472.6 & 664.4 \\
0.8 & 1.6  &  569.8 & 664.8 \\ 
0.9 & 1.7  &  618.5 & 665.0 \\
1.0 & 1.8  &  667.3 & 665.2 \\
1.2 & 2.0  &  764.8 & 665.7 \\
\hline
\end{tabular}
\end{table*}

\subsubsection{The $V-U$ and $e-J_{\rm z}$ diagrams}

\par To provide a deeper study of the kinematic properties
of CD-62$\degr$1346, we analyzed its position in the $V-U$
and $e-J_{\rm z}$ diagrams, where $e$ is the eccentricity and $J_{\rm
  z}$ the $Z$-component of the angular momentum, and also  examined
its orbit in space. In addition, as mentioned before, we also
investigated the kinematic behavior of other metal-poor barium and CH
stars.

\par Among the barium stars we selected three stars, HD~10613,
HD~123396 and HD~206983, which have already been investigated in
the literature and are also known to be metal-poor objects,
s-process enriched and have C/O ratios less than unity.  Among the
CH stars, we selected the early-type CH stars and Feh-Duf.
 We computed the distances for all  metal-poor barium stars and CH stars
as in Equation~(1), taking into consideration the temperature and
gravity already determined.  Bolometric corrections were also
determined from Alonso et al. (1999).  Interstellar absorption was
considered using the results from the literature, and is not higher
than $A_{\rm V}$\,=\,0.3.  Proper motions were also taken from
Hipparcos (Perryman 1997). We adopted
$M_{*}\,=\,0.8\,M_{\sun}$.  Table~9 shows the temperature and gravity
used to determine the distance, the radial velocities, the
Galactocentric space velocity components for the stars mentioned above,
and values of the velocity $V_{\rm GRF}$.  The last column of Table~9
gives the reference for the temperature, gravity, and radial velocity
for the studied stars.

\begin{table*} 
\caption{Atmospheric parameters (effective temperature and surface gravity), 
distance, radial velocity (RV), space velocity components relative 
to the Sun ($U_0,\, V_0$, $W_0$), and the Galactic rest frame velocity 
($V_{\rm GRF}$) for some metal-poor barium  and CH stars.} 
\begin{tabular}{ccccccccc}\hline
Star & $T_{\rm eff}$/log\,$g$ & Distance & RV & $U_0$  &  $V_0$  &  $W_0$  &  $V_{\rm GRF}$ & Ref.\\
      &  (K)/dex & (kpc) & km\,s$^{-1}$ & km\,s$^{-1}$  & km\,s$^{-1}$ & km\,s$^{-1}$ & 
km\,s$^{-1}$ &\\\hline
HD 10613   & 5100/2.8 & 0.40 & $+$89.3 & -155.0 &  150.2 & 128.4 & 251  & 1 \\  
HD 123396  & 4600/1.9 & 0.60 & $+$28.8 & -80.0  &  122.4 & 27.6  & 149  & 2 \\
HD 206983  & 4200/1.4 & 0.95 &  -319.2 & -67.2  & -147.2 & 162.5 & 229  & 3 \\
BD+04$\degr$2466 & 5100/1.8 & 1.70 & $+$38.5 & 95.2 & -242.0 & -215.8 & 338 & 1\\
CD-38$\degr$2151 & 4700/1.5 & 1.10 & $+$125.0 & -50.5 & 113.1 & -11.3 & 124 & 4\\
HD 26      & 5100/2.3 & 0.30 & -211.6  & -244.9 & -117.2 &  59.0  & 278 & 2 \\
HD 5223    & 4500/1.0 & 1.20 & -239.2  & -565.1 & -421.1 &  107.6 & 713 & 5 \\
HD 187861  & 4600/1.7 & 0.90 & $+$8.6  &  -72.2 &  -52.1 &  103.2 & 136 & 6 \\
HD 196944  & 5200/1.6 & 0.90 & -171.7  & -165.7 &   62.4 &  -89.8 & 199 & 2 \\
HD 198269  & 4800/1.3 & 0.80 & -198.6  &  -57.9 &  -10.9 & -138.5 & 151 & 7 \\
HD 201626  & 5200/2.3 & 0.30 & -149.4  &  -27.8 &   87.4 &   15.6 &  93 & 7 \\
HD 209621  & 4500/2.0 & 0.40 & -390.5  & -116.3 & -131.0 &  117.6 & 211 & 8\\
HD 224959  & 4900/2.2 & 0.65 & -130.5  &  256.9 &    5.5 &   26.3 & 258 & 6\\
Feh-Duf    & 4500/0.9 & 6.00 & $+$448.0 & -141.0 & -252.8 & -110.7 & 310 & 9\\\hline
\end{tabular}
\par
\par 1: $T_{\rm eff}$/log\,$g$/RV: Pereira \& Drake (2009);\,
\par 2: $T_{\rm eff}$/log\,$g$/RV: This work;\,
\par 3: $T_{\rm eff}$/log\,$g$/RV: Drake \& Pereira (2008);\,
\par 4: $T_{\rm eff}$/log\,$g$: Vanture (1992a),RV:\,Barbier-Brossat (1994);
\par 5: $T_{\rm eff}$/log\,$g$/RV: Goswami et al. (2006);\,
\par 6: $T_{\rm eff}$/log\,$g$: Masseron et al (2010),RV:Van Eck et al (2003);\,
\par 7: $T_{\rm eff}$/log\,$g$/RV: Van Eck et al. (2003);\,
\par 8: $T_{\rm eff}$/log\,$g$/RV: Goswami \& Aoki (2010);\,
\par 9: $T_{\rm eff}$/log\,$g$/RV: Drake \& Pereira (2011)\,
\end{table*}

\par Inspecting Table~9 we may conclude that among the metal-poor
barium stars, HD~206983 has the highest value of the $V_0$, similar
to the CH stars HD~26 and HD~209621. The high negative values, 
and those of some other CH stars, can be taken as an indication that they lag in the
Galactic rotation compared to the solar motion.
\\
However, the most interesting results arise from the space velocities
and $V_{\rm GRF}$ of the stars, BD+04$\degr$2466, HD 26 and HD 5223.
Excluding Feh-Duf, whose  kinematical behavior has already been
investigated (Drake \& Pereira 2011), these three objects present
the highest $V_{\rm GRF}$ among the CH stars.  The $V_{\rm GRF}$
value of HD~5223 and CD-62$\degr$1346 (569 km\,s$^{-1}$,
Section 4.3) is similar to, or even higher than, some hypervelocity stars
already investigated in the literature, such as
SDSS\,J130005.62\.+\,042201.6 (467 km\,s$^{-1}$, Tillich et al. 2010);
HD~271917 (615 km\,s$^{-1}$, Heber et al. 2008); 
HIP~60350 (535~km\,s$^{-1}$, Irrgang et al. 2010); and 
SDSS\,J013655.91\,+\,242546.0
(594~km\,s$^{-1}$, Tillich et al. 2009).  Figure~8 shows the position
of CD-62$\degr$1346, BD+04$\degr$2466, HD~5223 and Feh-Duf
and some hypervelocity stars in the $V$\,-\,$U$ diagram. The
kinematical data from the work of Holmberg et al. (2009) was used as a
reference.  In this work the authors determined the space velocities
and orbital parameters over 13\,000 nearby F and G dwarf stars.
CD-62$\degr$1346 and HD~5223 and hypervelocity stars lie very far
away from the Galactic disk population. Therefore, CD-62$\degr$1346
is a hypervelocity candidate because the value of $V_{\rm GRF}$ is
constrained by its mass, and hence its distance, while HD~5223 is a
hypervelocity star, not recognized before as such in the
literature. Its escape velocity according to the 
Galactic potential of Ortega et al. (2002) is 658~km\,s$^{-1}$.
These two stars are the first cool, chemically peculiar
stars to join the restricted group of hypervelocity B
stars first identified by Brown et al. (2005).

\par In the $e\,-\,J_{\rm z}$ diagram we show the position of
CD-62$\degr$1346, SDSS\,J130005.62\.+\,042201.6 (Tillich et al. 2010)
and SDSS\,121150.27$+$143762.2 (Tillich et al. 2011) and the F and G
dwarf stars from Holmberg et al. (2009). We do not show the position of
HD 5223 in this diagram since it has a $V_{\rm GRF}$ higher than the
escape velocity of the Galaxy.  Again, we see that CD-62$\degr$1346
lies very far away from the Galactic disk population.

\begin{figure} 
\includegraphics[width=8.5cm]{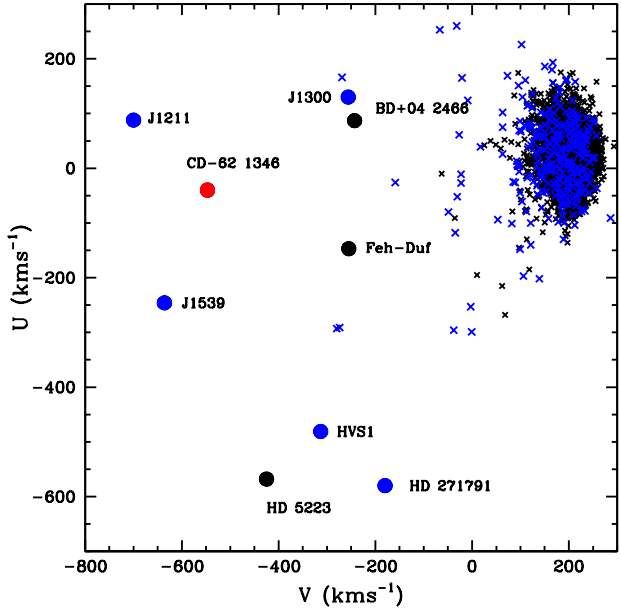}
 \caption{$V-U$ diagram for CD-62$\degr$1346 ({\sl red filled circle}) 
and other CH stars, BD+04$\degr$2466, HD~5223 and Feh-Duf
({\sl black filled circles}). 
We also show the positions of five hypervelocity stars ({\sl blue filled 
circles}).
The stars from Holmberg et al. (2009) serve as reference. Black 'x' points
represent stars with distances from the Galactic plane less than 1~kpc, while
blue 'x' points represent those above 1~kpc from the Galactic plane.}
\end{figure}

\begin{figure} 
\includegraphics[width=8.5cm]{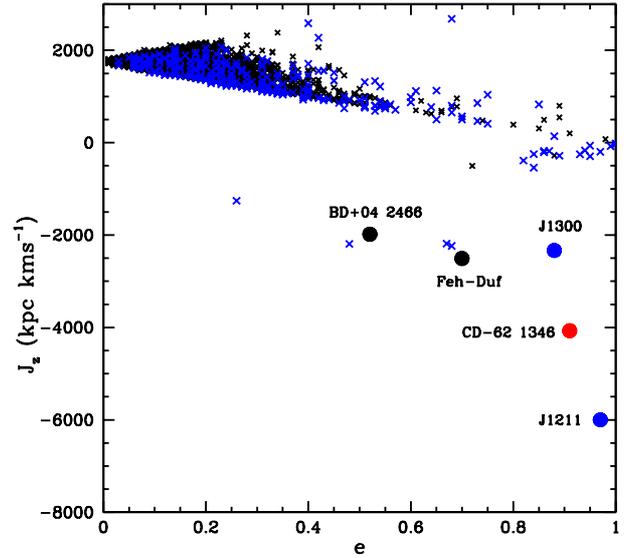}
\caption{CD-62$\degr$1346, BD+04$\degr$2466, Feh-Duf,
SDSS\,J130005.62\,+\,042201.6 and SDSS\,121150.27$+$14376.2
in the $e\,-\,J_{\rm z}$ diagram. Other symbols have the same meaning as in Figure~8.}
\end{figure}

\subsubsection{The dynamical orbit of BD+04$\degr$2466, CD-62$\degr$1346, 
HD~5223 and Feh-Duf in the Galaxy}

\par We also studied the dynamical time evolution of BD+04$\degr$2466,
CD-62$\degr$1346, HD~5223 and Feh-Duf in the Galaxy.  For this purpose
we computed their orbits integrating back in time beginning with the
present distance or initial $XYZ$ positions relative to the Sun and
with current spatial velocities $(UVW)$.  Figures~10 and 11 show the
past and future orbits of CD-62$\degr$1346 and HD~5223 for the adopted
present distances of 1.6~kpc and 1.2~kpc, respectively, and for the
time interval from $-$2.0 to $+$2.0~Gyr.  Figures~12 and 13 show the
past and future orbits of Feh-Duf and BD+04$\degr$2466 for the adopted
present distance of 6.0~kpc and 1.7~kpc, respectively, and for the
time interval from $-$2.0 Gyr to the present date.  These Figures show
the time evolution of the distances relative to  the Galactic Center,
which permits us to estimate the eccentricities of the orbits of
BD+04$\degr$2466, CD-62$\degr$1346 and Feh-Duf. The stars
BD+04$\degr$2466 and Feh-Duf have orbits typical of halo stars,
extending $\pm$10.0--20.0~kpc out of the plane of the Galactic
disk.  \\ CD-62$\degr$1346 has a highly eccentric orbit ($e$\,=\,0.9,
Figure~9) and is almost unbound to the Galaxy, traveling up to 50~kpc
from the plane of the Galactic disk and about 100~kpc from the
Galactic center. HD~5223, in turn, has an orbit that goes much
farther of the Galactic disk, $X\!=\! 600$~kpc (out of the range of
Figure~9), and its value from the Galactic plane ranges from $-$30 to
110~kpc for the same time interval from $-2$~Gyr to +2~Gyr. It is
interesting that CD-62$\degr$1346 and HD~5223 have basically the same
abundance pattern (low metallicity, carbon and s-process enrichment)
but are kinematically and dynamically different from the other CH
stars.  Because globular clusters are also known to have some CH stars
among its members, we cannot exclude the possibility that HD~5223
could have been ejected from one of these systems.  Alternatively,
HD~5223 could have an extragalactic origin and was captured by the
Milky Way after a tidal disruption of a nearby satellite galaxy and
will not belong to the Galaxy since its $V_{\rm GRF}$ exceeds the
Galaxy escape velocity.  CD-62$\degr$1346, like Feh-Duf (Drake \&
Pereira 2011), might has been captured by the Milky Way and became a
halo star (since it has a highly eccentric and retrograde orbit) or,
depending on the assumed mass and hence its distance,  it could be an
unbound star.

\begin{figure} 
\includegraphics[width=9.5cm]{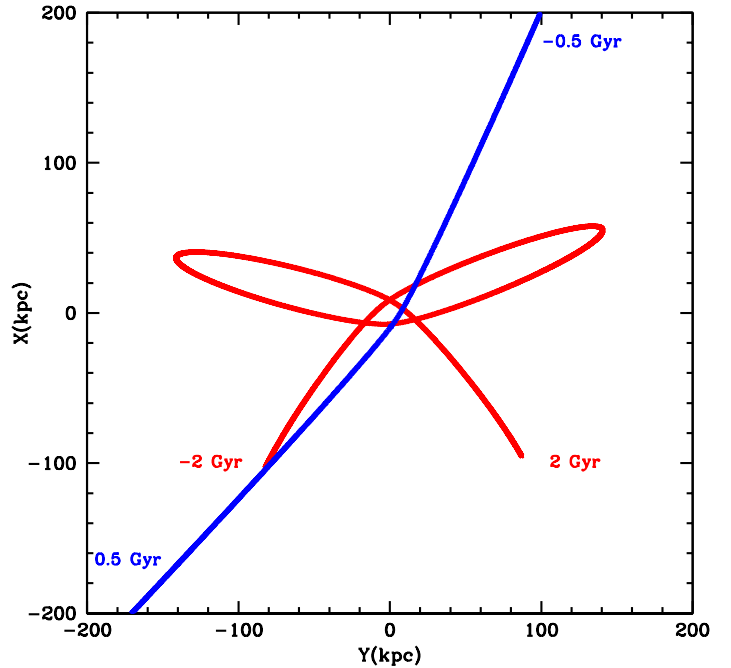}
\caption{Trajectory of CD-62$\degr$1346 (red) and HD~5223 (blue) in the $XY$ plane. 
For CD-62$\degr$1346 we show  the whole trajectory from $-2$~Gyrs to 2~Gyrs in time, 
while for HD~5223 we show only a part of it from $-0.5$~Gyr to 0.5~Gyr. The whole 
trajectory calculated from -2~Gyr to 2~Gyr extends from (X,Y)=(600,260) to 
(X,Y)=(-470,-430).}
\end{figure}

\begin{figure} 
\includegraphics[width=9.5cm]{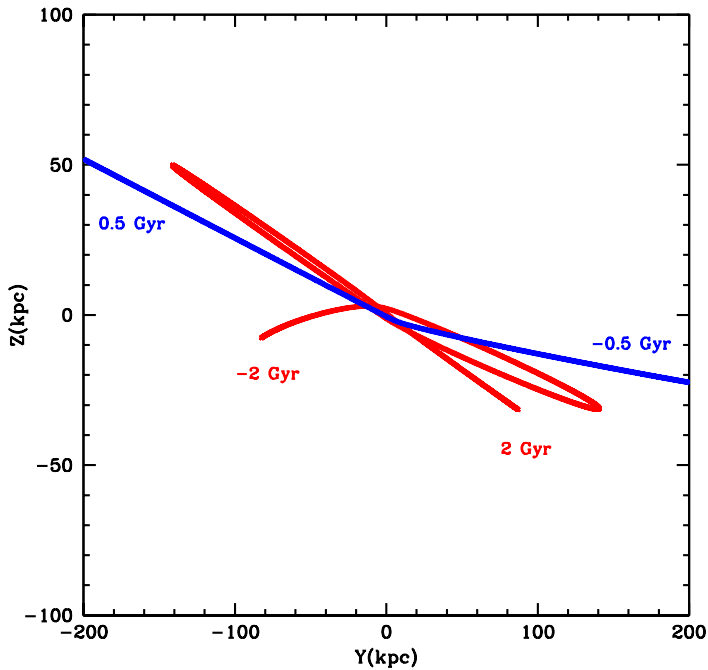}
\caption{Same as in Figure~10 but in the $YZ$ plane. For HD 5223 the whole trajectory 
extends from $(Z,Y)=(-30,260)$ to $(Z,Y)=(110,-400)$.}
\end{figure}

\begin{figure} 
\includegraphics[width=9.5cm]{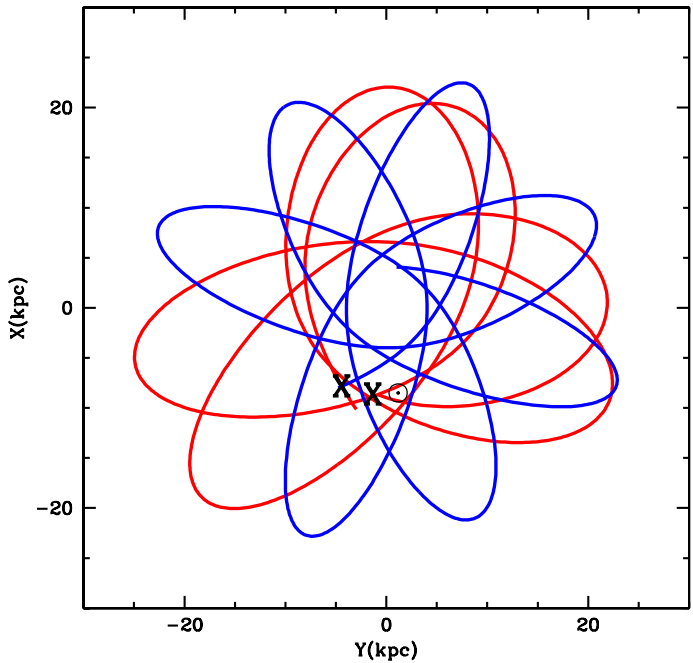}
\caption{Trajectory of BD+04$\degr$2466 (red) and Feh-Duf (blue) in the $XY$ plane. 
We show the orbits from $-2$~Gyr to the present position labeled with 'X'. We also show
the position of the Sun.}
\end{figure}

\begin{figure} 
\includegraphics[width=9.5cm]{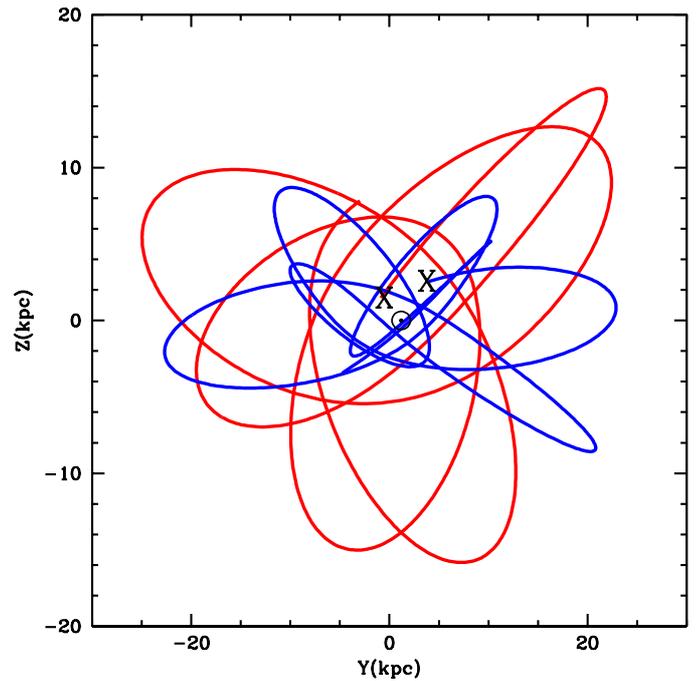}
\caption{Same as in Figure 12 but in the $YZ$ plane.}
\end{figure}

\section{Conclusions}

\par The results from our abundance and kinematical analysis performed by 
employing high-resolution optical spectra of the CH star CD-62$\degr$1346 can 
be summarized as follows :

\begin{enumerate}

\item CD-62$\degr$1346 is a CH star and also a ``lead star''
because its lead-to-cerium ratio is high ([Pb/Ce]\,=\,+0.80) and
closely follows the theoretical predictions for a star of this
metallicity. It is an evolved metal-deficient
giant with $T_{\rm eff}=5300$~K and $\log g=1.7$. We found also that
this star has the high carbon ([C/Fe]) and s-element abundance typical for
CH stars.

\item CD-62$\degr$1346 has to be excluded from the Kapteyn group
(Wylie de Boer et al. 2010) since the distance adopted by these
authors is too small.  In addition, the position in the
$e\,-\,J_{\rm z}$ diagram shows that CD-62$\degr$1346 has an extreme
retrograde motion, high eccentricity, and very negative angular
momentum.  The position of CD-62$\degr$1346 in $J_{\perp}-J_{\rm z}$
diagram of Kepley et al. (2007) $(1446,-4072)$ clearly indicates
that it belongs to the halo population.

\item Detailed kinematic analysis showed that CD-62$\degr$1346 has a
$V_{\rm GRF}$ close to the Galaxy escape velocity, making it marginally bound. 
This is constrained by the adopted mass and hence the distance.  The extreme
retrograde motion may suggest that the CD-62$\degr$1346 has an
extragalactic origin and was captured by the Milky Way.  However,
the high $\alpha$-element abundances, typical for halo stars of this
metallicity, do not support this suggestion.  Nissen \& Schuster
(2010) showed that dwarf stars could be distinguished according to
their [$\alpha$/Fe] ratios. Those with high [$\alpha$/Fe] ratios
would belong to the halo population and those with low [$\alpha$/Fe]
ratios may have been captured from nearby galaxies.  Feh-Duf, like
other extreme retrograde stars with low $\alpha$-element abundances,
would be another example (Venn et al. 2003, Drake \& Pereira 2011).

\item Among the CH stars, we found that HD~5223 is not bound to the Galaxy
because its $V_{\rm GRF}$ exceeds the Galaxy escape velocity. 
It is unbound for any Galactic potential considered in this work.
Tidal disruption of a satellite galaxy or ejection from a globular
cluster are good suggestions to explain its dynamical orbit through the Galaxy.

\item Among other objects that display the same abundance pattern,
s-process enrichment, halo membership, and binarity, the
yellow symbiotics (Pereira \& Roig 2009) were also kinematically
investigated to check whether they would behave like some of these
CH stars. Although their distances are more difficult to constrain
due to, in some cases, to uncertainties in the interstellar absorption, none
of them displays high $V_{\rm GRF}$ or has retrograde
motion. CEMP-s stars are another group that would deserve
investigation because they display the same phenomena, mass-transfer
with s-process enrichment in a binary system. Indeed, this has
already been done by Carollo et al. (2012). Although they found many stars
with significant retrograde evidence ($V_{\phi}$ $<$ --200~km\,s$^{-1}$), 
none of them seems to have $V_{\rm GRF}$ higher than
the Galaxy escape velocity.

\item Hypervelocity stars are good candidates to constraint the
Galactic potential (Kenyon et al. 2008). CD-62$\degr$1346 is another
example of a star that may be bound or unbound according to the Galactic
potential adopted. CD-62$\degr$1346 and HD 5223 are the first
red giants to join the restricted group of hypervelocity stars which
formerly consisted of B-type stars only.

\end{enumerate}

\begin{acknowledgements} 
N.A.D. acknowledges support of the Saint Petersburg State University,
Russia, under the Project 6.38.73.2011. 
We also acknowledge the use of electronic databases (VALD, SIMBAD, NASA's ADS).
\end{acknowledgements}

\Online

\longtab{1}{
\begin{longtable}{cccccc}
\caption{Observed Fe\,{\sc i} and Fe\,{\sc ii} lines. The $\log gf$ values 
were taken from Lambert et al. (1996). }\\
\hline\hline
Element &  $\lambda$\,(\AA) &  $\chi$(eV) & $\log gf$ & $W_\lambda$ (m\AA) \\\hline
Fe\,{\sc i} & 5133.60  &   4.18  & 0.20 & 78 \\  
&  5151.91  &    1.01  &  -3.32  &  74 \\    
&  5159.06  &    4.28  &  -0.65  &  23 \\    
&  5162.27  &    4.18  &   0.07  &  65 \\    
&  5171.60  &    1.49  &  -1.76  & 119 \\
&  5194.94  &    1.56  &  -2.09  & 104 \\    
&  5232.94  &    2.94  &  -0.08  & 125 \\
&  5242.49  &    3.63  &  -0.97  &  38 \\    
&  5250.21  &    0.12  &  -4.92  &  30 \\    
&  5281.79  &    3.04  &  -0.83  &  81 \\    
&  5288.52  &    3.69  &  -1.51  &  12 \\    
&  5307.36  &    1.61  &  -2.97  &  44 \\    
&  5322.04  &    2.28  &  -2.84  &  17 \\    
&  5339.93  &    3.27  &  -0.68  &  75 \\    
&  5341.02  &    1.61  &  -1.95  & 103 \\
&  5364.87  &    4.45  &   0.23  &  54 \\    
&  5367.47  &    4.42  &   0.44  &  67 \\    
&  5369.96  &    4.37  &   0.54  &  70 \\    
&  5389.48  &    4.42  &  -0.25  &  26 \\    
&  5400.50  &    4.37  &  -0.10  &  41 \\
&  5405.77  &    0.99  &  -1.85  & 140 \\
&  5410.91  &    4.47  &   0.40  &  60 \\
&  5434.52  &    1.01  &  -2.12  & 126 \\
&  5445.04  &    4.39  &   0.04  &  48 \\    
&  5487.75  &    4.32  &  -0.65  &  22 \\
&  5506.78  &    0.99  &  -2.80  & 100 \\    
&  5554.90  &    4.55  &  -0.38  &  21 \\
&  5638.26  &    4.22  &  -0.72  &  20 \\    
&  5658.82  &    3.40  &  -0.81  &  63 \\ 
&  5686.53  &    4.55  &  -0.45  &  15 \\ 
&  5762.99  &    4.21  &  -0.41  &  40 \\
&  5883.82  &    3.96  &  -1.21  &  15 \\   
&  5934.65  &    3.93  &  -1.02  &  19 \\    
&  6024.06  &    4.55  &  -0.06  &  39 \\    
&  6027.05  &    4.08  &  -1.09  &  13 \\    
&  6056.01  &    4.73  &  -0.40  &  16 \\    
&  6065.48  &    2.61  &  -1.53  &  72 \\    
&  6136.61  &    2.45  &  -1.40  &  85 \\    
&  6137.69  &    2.59  &  -1.40  &  82 \\    
&  6200.31  &    2.60  &  -2.44  &  20 \\   
&  6213.43  &    2.22  &  -2.48  &  34 \\    
&  6252.56  &    2.40  &  -1.72  &  74 \\    
&  6265.13  &    2.18  &  -2.55  &  38 \\     
&  6393.60  &    2.43  &  -1.43  &  82 \\    
&  6430.85  &    2.18  &  -2.01  &  72 \\
&  6430.85  &    2.18  &  -2.01  &  72 \\
Fe\,{\sc ii} & 4993.35 &   2.81  & -3.67 & 30 \\  
&   5197.56  &   2.81  &  -2.25  & 90 \\
&   5325.56  &   3.22  &  -3.17  & 37 \\  
&   5414.05  &   3.22  &  -3.62  & 19 \\  
&   5425.25  &   3.20  &  -3.21  & 32 \\  
&   5534.83  &   3.25  &  -2.77  & 52 \\  
&   5991.37  &   3.15  &  -3.56  & 18 \\  
&   6084.10  &   3.20  &  -3.80  & 18 \\  
&   6149.25  &   3.89  &  -2.72  & 27 \\  
&   6247.55  &   3.89  &  -2.34  & 40 \\  
&   6416.92  &   3.89  &  -2.68  & 22 \\  
&   6432.68  &   2.89  &  -3.58  & 33 \\\hline
\end{longtable}
}

\longtab{3}{
\begin{longtable}{ccccccc}
\caption{Other lines studied}\\
\hline\hline
$\lambda$\,(\AA) & Element & $\chi$(eV) & $\log gf$ & Ref & $EW_\lambda$ (m\AA) \\
\hline

5682.65  & Na\,{\sc i}  & 2.10 &   -0.700 &  PS &  12\\
5688.22  & Na\,{\sc i}  & 2.10 &   -0.400 &  PS &  26\\
4730.04  &  Mg\,{\sc i} & 4.34 &   -2.390 &  R03  &   21\\
5528.42  &  Mg\,{\sc i} & 4.34 &   -0.490 &  JJ   &  133\\
5711.10  &  Mg\,{\sc i} & 4.34 &   -1.750 &  R99  &   38\\
6319.24  &  Mg\,{\sc i} & 5.11 &   -2.160 &  Ca07 &   12\\
7387.70  &  Mg\,{\sc i} & 5.75 &   -0.870 &  MR94 &   25\\
6125.03  &  Si\,{\sc i} & 5.61 &   -1.540 &  E93 &   10\\
6145.02  &  Si\,{\sc i} & 5.61 &   -1.430 &  E93 &   10\\
6155.14  &  Si\,{\sc i} & 5.62 &   -0.770 &  E93 &   25\\
7800.00  &  Si\,{\sc i} & 6.18 &   -0.720 &  E93 &   10\\
8728.01  &  Si\,{\sc i} & 6.18 &   -0.360 &  E93 &   32\\
8742.45  &  Si\,{\sc i} & 5.87 &   -0.510 &  E93 &   46\\
6102.73  &  Ca\,{\sc i} & 1.88 &   -0.790 &  D2002 &   78\\
6122.23  &  Ca\,{\sc i} & 1.89 &   -0.320 &  D2002 &  116\\
6161.30  &  Ca\,{\sc i} & 2.52 &   -1.270 &  E93   &   28\\
6162.18  &  Ca\,{\sc i} & 1.90 &   -0.090 &  D2002 &  134\\
6166.44  &  Ca\,{\sc i} & 2.52 &   -1.140 &  R03   &   22\\
6169.04  &  Ca\,{\sc i} & 2.52 &   -0.800 &  R03   &   35\\
6169.56  &  Ca\,{\sc i} & 2.53 &   -0.480 &  DS91  &   49\\
6439.08  &  Ca\,{\sc i} & 2.52 &    0.470 &  D2002 &  113\\
6455.60  &  Ca\,{\sc i} & 2.51 &   -1.290 &  R03   &   18\\
6471.66  &  Ca\,{\sc i} & 2.51 &   -0.690 &  S86   &   46\\
6493.79  &  Ca\,{\sc i} & 2.52 &   -0.110 &  DS91  &   79\\
4512.74  &  Ti\,{\sc i} &  0.84 &   -0.480 &  MFK &   31\\
4518.03  &  Ti\,{\sc i} &  0.83 &   -0.320 &  MFK &   39\\
4533.25  &  Ti\,{\sc i} &  0.85 &    0.480 &  MFK &   84\\
4534.78  &  Ti\,{\sc i} &  0.84 &    0.281 &  MFK &   69\\
4548.77  &  Ti\,{\sc i} &  0.83 &   -0.350 &  MFK &   36\\
4617.28  &  Ti\,{\sc i} &  1.75 &    0.389 &  MFK &   26\\
4681.92  &  Ti\,{\sc i} &  0.05 &   -1.070 &  MFK &   41\\
4758.12  &  Ti\,{\sc i} &  2.25 &    0.425 &  MFK &   14\\
4981.74  &  Ti\,{\sc i} &  0.85 &    0.500 &  MFK &   93\\
5016.17  &  Ti\,{\sc i} &  0.85 &   -0.574 &  MFK &   26\\
5022.87  &  Ti\,{\sc i} &  0.83 &   -0.434 &  MFK &   49\\
5039.96  &  Ti\,{\sc i} &  0.02 &   -1.130 &  MFK &   48\\
5145.47  &  Ti\,{\sc i} &  1.46 &   -0.574 &  MFK &   10\\
5152.19  &  Ti\,{\sc i} &  0.02 &   -2.024 &  MFK &   16\\
5173.75  &  Ti\,{\sc i} &  0.00 &   -1.118 &  MFK &   46\\
5210.39  &  Ti\,{\sc i} &  0.05 &   -0.883 &  MFK &   61\\
6258.11  &  Ti\,{\sc i} &  1.44 &   -0.355 &  MFK &   25\\
5247.57  &  Cr\,{\sc i} &  0.96 &   -1.630 &  MFK &   25\\
5296.70  &  Cr\,{\sc i} &  0.98 &   -1.390 &  GS  &   36\\
5298.28  &  Cr\,{\sc i} &  0.98 &   -1.160 &  MFK &   57\\
5300.75  &  Cr\,{\sc i} &  0.98 &   -2.130 &  GS  &   12\\
5345.81  &  Cr\,{\sc i} &  1.00 &   -0.980 &  GS  &   64\\
5409.79  &  Cr\,{\sc i} &  1.03 &   -0.720 &  GS  &   74\\
4904.42  &  Ni\,{\sc i} &  3.54 &   -0.170 &  MFK &   35\\
4913.98  &  Ni\,{\sc i} &  3.74 &   -0.600 &  MFK &   19\\
4935.83  &  Ni\,{\sc i} &  3.94 &   -0.360 &  MFK &   15\\
5010.94  &  Ni\,{\sc i} &  3.63 &   -0.870 &  MFK &   12\\
5578.73  &  Ni\,{\sc i} &  1.68 &   -2.640 &  MFK &   17\\
6108.12  &  Ni\,{\sc i} &  1.68 &   -2.440 &  MFK &   16\\
6176.82  &  Ni\,{\sc i} &  4.09 &   -0.264 &  MFK &   13\\
6586.32  &  Ni\,{\sc i} &  1.95 &   -2.810 &  MFK &   10\\
6767.78  &  Ni\,{\sc i} &  1.83 &   -2.170 &  MFK &   29\\
4883.68  &  Y\,{\sc ii} &  1.08 &    0.070 & SN96 &  101\\
5087.43  &  Y\,{\sc ii} &  1.08 &   -0.170 & SN96 &   81\\
5123.21  &  Y\,{\sc ii} &  0.99 &   -0.930 & SN96 &   50\\
5205.72  &  Y\,{\sc ii} &  1.03 &   -0.340 & SN96 &   83\\
5402.78  &  Y\,{\sc ii} &  1.84 &   -0.440 & R03  &   26\\
4208.99  &  Zr\,{\sc ii} &  0.71 &  -0.460 & SN96 &  91\\
4317.32  &  Zr\,{\sc ii} &  0.71 &  -1.380 & SN96 &  48\\
4496.97  &  Zr\,{\sc ii} &  0.71 &  -0.590 & SN96 & 108\\
5112.27  &  Zr\,{\sc ii} &  1.66 &  -0.760 & SN96 &  21\\
5477.82  &  Zr\,{\sc ii} &  1.83 &  -1.400 & SN96 &  10\\
6496.90  &  Ba\,{\sc ii} &  0.60 &  -0.380 & WM80 & 222\\
4934.83  &  La\,{\sc ii} &  1.25 &   -0.920 & VWR  &  14\\
5122.99  &  La\,{\sc ii} &  0.32 &   -0.930 & SN96 &  54\\
5303.53  &  La\,{\sc ii} &  0.32 &   -1.350 & VWR  &  30\\
6320.43  &  La\,{\sc ii} &  0.17 &   -1.520 & SN96 &  32\\
6390.48  &  La\,{\sc ii} &  0.32 &   -1.410 & VWR  &  31\\
4120.84  &  Ce\,{\sc ii} &  0.32 &   -0.240 &  SN96 & 71\\
4127.38  &  Ce\,{\sc ii} &  0.68 &    0.240 &  SN96 & 72\\
4222.60  &  Ce\,{\sc ii} &  0.12 &   -0.180 &  SN96 & 82\\
4418.79  &  Ce\,{\sc ii} &  0.86 &    0.310 &  SN96 & 71\\
4486.91  &  Ce\,{\sc ii} &  0.30 &   -0.360 &  SN96 & 77\\
4562.37  &  Ce\,{\sc ii} &  0.48 &    0.330 &  SN96 & 90\\
4628.16  &  Ce\,{\sc ii} &  0.52 &    0.260 &  SN96 & 89\\
5117.17  &  Ce\,{\sc ii} &  1.40 &    0.010 &  VWR  & 18\\
5187.46  &  Ce\,{\sc ii} &  1.21 &    0.300 &  VWR  & 44\\
5274.24  &  Ce\,{\sc ii} &  1.28 &    0.389 &  VWR  & 49\\
5409.23  &  Ce\,{\sc ii} &  1.10 &   -0.375 &  VWR  & 33\\
5472.30  &  Ce\,{\sc ii} &  1.25 &   -0.190 &  VWR  & 22\\
6051.80  &  Ce\,{\sc ii} &  0.23 &   -1.600 &  S96  & 11\\
4811.34  & Nd\,{\sc ii}  &  0.06 &   -1.015 &  VWR  & 59\\
4820.34  & Nd\,{\sc ii}  &  0.20 &   -1.161 &  VWR  & 50\\
4959.12  & Nd\,{\sc ii}  &  0.06 &   -0.916 &  VWR  & 64\\
4989.95  & Nd\,{\sc ii}  &  0.63 &   -0.624 &  VWR  & 53\\
5063.72  & Nd\,{\sc ii}  &  0.98 &   -0.758 &  VWR  & 23\\
5089.83  & Nd\,{\sc ii}  &  0.20 &   -1.140 &  E93  & 28\\
5092.80  & Nd\,{\sc ii}  &  0.38 &   -0.510 &  E93  & 49\\
5130.59  & Nd\,{\sc ii}  &  1.30 &    0.100 &  SN96 & 49\\
5234.19  & Nd\,{\sc ii}  &  0.55 &   -0.460 &  SN96 & 65\\
5249.58  & Nd\,{\sc ii}  &  0.98 &    0.080 &  SN96 & 68\\
5293.16  & Nd\,{\sc ii}  &  0.82 &   -0.200 &  SN96 & 62\\
5311.46  & Nd\,{\sc ii}  &  0.98 &   -0.560 &  SN96 & 27\\
5319.81  & Nd\,{\sc ii}  &  0.55 &   -0.350 &  SN96 & 70\\
5361.47  & Nd\,{\sc ii}  &  0.68 &   -0.400 &  SN96 & 47\\
5431.54  & Nd\,{\sc ii}  &  1.12 &   -0.457 &  VWR  & 18\\
5442.26  & Nd\,{\sc ii}  &  0.68 &   -0.900 &  SN96 & 17\\
5740.88  & Nd\,{\sc ii}  &  1.16 &   -0.560 &  VWR  & 15\\\hline
\hline
\footnote{References to Table 4:
\par Ca07: Carretta et al. (2007);
\par D2002: Depagne et al. (2002;)
\par DS91: Drake \& Smith (1991);
\par E93: Edvardsson et al. (1993);
\par GS: Gratton \& Sneden (1988);
\par PS: Preston \& Sneden (2001);
\par R03: Reddy et al. (2003); 
\par MFK: Martin et al. (2002);
\par MR94: McWilliam \& Rich (1994);
\par S86: Smith et al. (1986);
\par S96: Smith et al. (1996);
\par SN96: Sneden et al. (1996); 
\par VWR: van Winckel \& Reyniers (2000);
\par WM80: Wiese \& Martin (1980);
}\\
\end{longtable}
}

\end{document}